\documentclass[useAMS,usenatbib]{mn2e}

\usepackage[dvips]{graphicx,rotating}

\newcommand{\nh}{$N_{\mathrm{H}}$}
\newcommand{\z}{\phantom{0}}
\newcommand{\rxte}{\textit{RXTE}}
\newcommand{\inte}{\textit{INTEGRAL}}
\newcommand{\igr}{IGR~J19140$+$0951}

\title[Peering through the stellar wind of \igr...]{Peering through the stellar wind of \igr\ with simultaneous \inte /\rxte\ observations}
\author[L. Prat, J. Rodriguez, D.C. Hannikainen and S.E. Shaw]{L. Prat$^{1}$\thanks{E-mail:
lionel.prat@cea.fr}, J. Rodriguez$^{1}$, D.C. Hannikainen$^{2,3}$ and S.E. Shaw$^4$\\
$^{1}$DSM/IRFU/Service d'Astrophysique, CEA/Saclay, F-91191 Gif-sur-Yvette, France\\
$^{2}$Observatory, PO Box 14, FI-00014 University of Helsinki, Finland\\
$^{3}$Mets\"ahovi Radio Observatory, Helsinki University of Technology TKK, Mets\"ahovintie 114, FI-02540 Kylm\"al\"a, Finland\\
$^{4}$School of Physics and Astronomy, University of Southampton, S017 1BJ, UK}
\begin{document}

\date{Accepted . Received ; in original form}

\pagerange{\pageref{firstpage}--\pageref{lastpage}} \pubyear{2007}

\maketitle

\label{firstpage}

\begin{abstract}
We have used the \rxte\ and \inte\ satellites simultaneously to observe the High Mass X-ray binary \igr. The spectra obtained in the 3--80 keV range have allowed us to perform a precise spectral analysis of the system along its binary orbit. The spectral evolution confirms the supergiant nature of the companion star and the neutron star nature of the compact object. Using a simple stellar wind model to describe the evolution of the photoelectric absorption, we were able to restrict the orbital inclination angle in the range 38\degr--75\degr. This analysis leads to a wind mass-loss rate from the companion star of $\sim$ 5$\times$ 10$^{-8}$ M$_{\sun}$/year, consistent with an OB I spectral type. We have detected a soft excess in at least four observations, for the first time for this source. Such soft excesses have been reported in several HMXBs in the past. We discuss the possible origin of this excess, and suggest, based on its spectral properties and occurrences around the superior conjunction, that it may be explained as the reprocessing of the X-ray emission originating from the neutron star by the surrounding ionised gas.

\end{abstract}


\section{Introduction}

High Mass X-ray Binaries (HMXBs) are binary systems consisting of a compact object orbiting a massive companion star. Prior to the launch of the INTErnational Gamma-ray Astrophysics Laboratory (\inte) in 2002, a large majority of the known HMXBs contained a Be companion. In Be-type HMXBs, the compact object emits strong X-ray flashes when it crosses the equatorial plane of the companion star, where a thick disk of matter originating from the stellar wind is present. Supergiant O and B-type stars have more isotropic stellar winds which absorb the X-ray emission of the compact object, rendering them almost undetectable below a few keV. Thanks to its sensitivity in the soft gamma-ray range, however, \inte\ has found many such systems in the past few years \citep[see e.g.][]{Catalogue, Catalogue_tout}. In this perspective, the use of X-ray spectroscopy at different orbital phases makes it possible to probe the stellar wind, providing two-dimensional information on the density and ionization structure of the wind. For instance, the soft excess that is present in the soft X-ray spectra of many HMXBs, whose origin is still quite mysterious, is linked to the physics of the wind close to the compact object, especially the region where the fast moving stellar wind collides with the slow moving and highly ionized gas surrounding the compact object \citep{soft}.

\igr\ was discovered by \inte\ in March 2003 \citep{decouverte}, during the first observation of the nearby microquasar GRS 1915+105\footnote{\igr\ lies $\sim 1.1$\degr\ from GRS 1915+105.} \citep[][]{decouvre}. \citet{periode2} and \citet{periode} analysed archival \rxte /All-Sky Monitor (ASM) data from 1996 onwards and discovered a 13.552 $\pm$ 0.006 day periodicity in the X-ray light curve. They associated it with the orbital period of the system, identified as an X-ray binary. An early analysis of the first \rxte\ and \inte\ data is presented in \citet[][hereafter Paper I]{Jerome}. These observations were not simultaneous, but the study of the spectral evolution of the source allowed them to classify it as a probable HMXB hosting a neutron star, which captures matter from the stellar wind without Roche lobe overflow. In this paper, we confirm this classification for Ê\igr. There is no evidence in the \rxte\ data of an eclipse of the compact objet by its companion star, which gives an upper limit to the inclination of the system (see Sec. \ref{model}). After refinement of the X-ray position with \textit{Chandra}, \citet{optique} identified the infrared counterpart with 2MASS 19140422+0952577. \citet{nespoli07} later estimated the spectral type of the companion star to be B1 I. \citet{Hanni} made further observations, narrowing the spectral type to B0.5 I, confirming the HMXB nature of \igr. \igr\ is situated in the direction of the Sagittarius arm, a site of active stellar formation. Given that HMXBs are young systems and that the Sagittarius arm already hosts several HMXBs, it is reasonable to think that \igr's distance lies between 2 and 6 kpc. This is further supported by the upper limit of 5~kpc obtained from the K-magnitudes of the source \citep{Hanni}.

In this paper, we present several \rxte\ and \inte\ observations of \igr, most of them taken simultaneously with the two satellites. The sequence of observations and data reduction process is described in Section \ref{data}. In Section \ref{resultats}, we present the spectral analysis and its main results. In Section \ref{discut}, we discuss these results and use a simple stellar wind model to calculate some characteristics of the source.

\begin{table*}
 \centering
 \begin{minipage}{140mm}
  \caption{List of the observations.}
  \label{logg}
		\begin{tabular}{@{\extracolsep{20pt}}ccccc}
  \hline
			Observation ID & \inte      & Date\footnote{Midpoint of observations}  & \rxte\ Exposure & Orbital %
			Phase\footnote{The orbital phases are calculated using an orbital period of 13.552 days, and a phase 0.5 in MJD 51593.4 \citep{periode2}} \\
			               & revolution & (MJD)                                    &   (s)    &    (cycles)   \\
 \hline
		P80404-01-01-00 & 0049 & 52,708.847 & 2864 & 0.815 \\
		P90112-01-01-00 &      & 53,087.593 & 7760 & 0.763 \\
		P90112-01-02-00 & 0246 & 53,296.931 & 8752 & 0.210 \\
		P90112-01-03-00 & 0295 & 53,443.185 & 1936 & 0.002 \\
		P90112-02-01-00 & 0305 & 53,473.889 & 5808 & 0.268 \\
		P90112-02-01-01 & 0305 & 53,473.202 & 3232 & 0.217 \\
		P90112-02-02-00 & 0361 & 53,641.355 & 2016 & 0.625 \\
		P90112-02-03-00 &      & 53,676.250 & 3216 & 0.199 \\
		P90112-03-01-00 & 0373 & 53,677.239 & 9152 & 0.274 \\
		P90112-03-02-00 & 0431 & 53,852.034 & 2352 & 0.170 \\
		P91083-01-01-00 & 0478 & 53,990.140 & 1776 & 0.361 \\
		P91083-01-01-01 & 0478 & 53,990.207 & 2176 & 0.366 \\
		P91083-01-02-00 & 0480 & 53,996.118 & 2416 & 0.803 \\
		P91083-01-03-00 & 0481 & 54,001.625 & 2272 & 0.209 \\
		P91083-01-04-00 & 0486 & 54,016.558 & 1488 & 0.311 \\
		P91083-01-05-00 & 0489 & 54,025.374 & 2176 & 0.961 \\
		P91083-01-06-00 & 0493 & 54,037.379 & 1408 & 0.847 \\
		P91083-01-06-01 & 0495 & 54,040.956 & 1520 & 0.111 \\
		P91083-01-07-00 &      & 54,049.996 & 4816 & 0.778 \\
		P91083-02-01-00 & 0501 & 54,059.039 & 3200 & 0.445 \\
		P91083-02-02-00 & 0537 & 54,166.928 & 1680 & 0.406 \\
		P91083-02-03-00 & 0539 & 54,173.148 & 1520 & 0.865 \\
		P91083-02-04-00 & 0542 & 54,181.948 & 4336 & 0.514 \\
		P91083-02-05-00 & 0544 & 54,187.806 & 6064 & 0.947 \\
		P91083-02-06-00 & 0546 & 54,195.604 & 1904 & 0.523 \\
		P91083-02-06-01 & 0546 & 54,195.537 & 1504 & 0.518 \\
		P91083-03-01-00 & 0549 & 54,202.729 & 6832 & 0.048 \\
		P91083-03-02-00 & 0551 & 54,208.841 & 3216 & 0.499 \\
		P91083-03-03-00 & 0553 & 54,214.537 & 1920 & 0.920 \\
		P91083-03-04-00 & 0557 & 54,226.594 & 3040 & 0.809 \\
		P91083-03-05-00 & 0559 & 54,232.526 & 2160 & 0.247 \\
		P91083-03-05-01 & 0559 & 54,232.594 & 2528 & 0.252 \\
		P91083-04-01-01 &      & 54,373.323 & 2720 & 0.636 \\
		Chandra         &      & 53,136.732 &  & 0.387 \\
\hline
\end{tabular}
\end{minipage}
\end{table*}

\section{Observations and data reduction}
\label{data}

We have been monitoring GRS 1915+105 (and therefore \igr) since March 2003 with \inte, and made simultaneous \rxte\ observations as often as possible. Our data cover the period March 2004--September 2007 with 32 observations, the log of which is reported in Table \ref{logg}. We also used one public \rxte\ observation performed in March 2003, and used the results from \citet{optique} based on one \textit{Chandra} observation taken in May 2004. 

\subsection{\inte/IBIS data}
To study the properties of the hard ($>$ 20 keV) X-ray emission from the source, we used the data from the IBIS telescope on board \inte\ \citep{IBIS_det}. This instrument uses a coded mask, which allows imaging over a large Field of View (FOV), $\sim$ 30\degr$\times$30\degr{\z}up to zero response. The Totally Coded FOV (TCFOV), where the sensibility is uniform, is 9\degr$\times$9\degr. The \inte\ Soft Gamma-Ray Imager \citep[ISGRI,][]{ISGRI_det} is the upper layer of the IBIS detection unit, covering the range between 13 keV and a few hundred keV. Its angular resolution is 12\arcmin.

The \inte\ campaign was aimed at studying the microquasar GRS 1915+105, but also allows observations of every other source that lies in the IBIS TCFOV, including \igr. Since October 2004, our \rxte\ observations were made simultaneously with the \inte\ observations. Following the method used in Paper I, the data were reduced using the standard {\tt{Off-line Scientific Analysis (OSA)}} v. 7.0 software package provided by the \inte\ Science Data Centre (http://isdc.unige.ch). First we ran the software up to the production of images and mosaics in the 20--40 and 40--80 keV range, using only the science windows simultaneous with the \rxte\ observations. The software was left free to find the most significant sources in the images, which means that faint sources like \igr\ were not always detected. Then we used a catalogue containing the 7 most luminous sources of the field for spectral extraction, in order to ``force" the program to extract the spectrum of our source from the images. Given the faintness of the source, we rebinned the redistribution matrix file provided with OSA to obtain 6 energy bins in the range 18--110 keV.

\begin{figure*}
	\includegraphics[width=17.5cm]{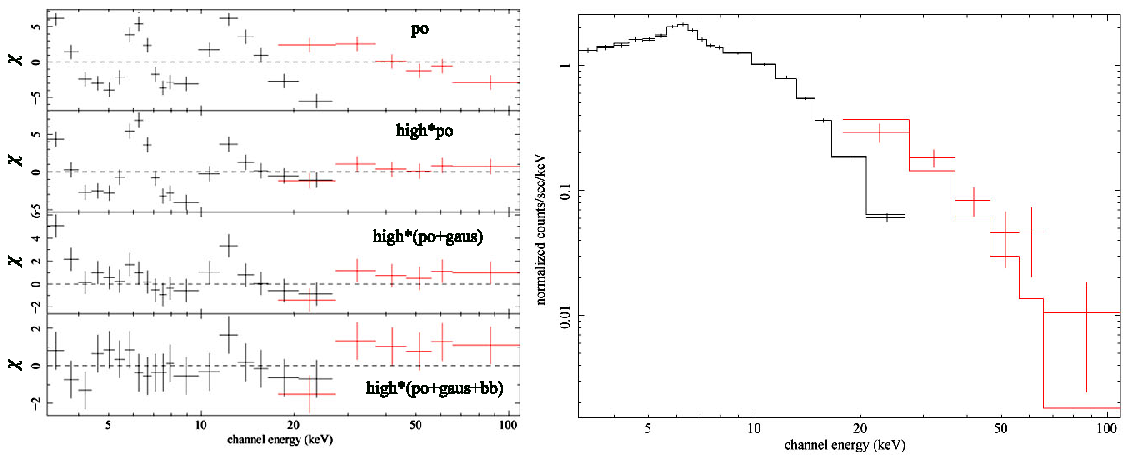}
	\caption{\textbf{Left: } Residuals in terms of $\sigma$ between the models used to fit the \inte +\rxte\ spectra. In each case the model mentioned in the panel is convolved with interstellar absorption. Po stands for power-law, high stands for high-energy cut-off, gaus stands for Gaussian and bb stands for black body. \textbf{Right: } Joint \inte +\rxte\ spectrum from ObsID P91083-03-02-00, corresponding to the residuals on the left. \rxte/PCA data are from 3--25 keV and \inte/ISGRI data from 20--100 keV. The best model is superposed on the spectrum.}
	\label{exemple}
\end{figure*}

\subsection{\rxte/PCA data}

For our analysis of the \rxte\ data, we used the top layers of the Proportional Counter Units (PCUs) 2 and 3 of the Proportional Counter Array (PCA), which were turned on in every observation. The data were reduced using {\tt{LHEASOFT}} package v6.4. We restricted the study to the time when the elevation angle was above 10\degr\ and the offset pointing was less than 0\fdg02, as recommended by the GOF (Guest Observer Facility) to avoid Earth contamination and errors due to slew motion. In addition, and since the source is quite weak, we further rejected times of high electron background in the PCA (i.e. times when the electron ratio in one of the PCUs is greater than 0.1). The response and background were generated using  {\tt{PCARSP}} version 10.1, and {\tt{PCABACKEST}} version 3.1 respectively, with the latest calibration files provided by the \rxte\ GOF. These files include the corrections applied to the PCA Background since 2007 September 18. We verified for \igr\ that the impact of these corrections on the background is typically 5\% below 5 keV, and 2\% above, which is what is expected for a faint source. The source is faint and the errors dominated by statistical effects \citep{Jahoda}, so we did not include any systematic error. We however checked that the results remained the same when adding a ``standard" systematic error of 0.6\% to the data, with only a slight and non significant decrease in the reduced $\chi^2$. In addition, the default spectral binning of the PCA detector is too fine for this dim source, so we rebinned all our \rxte/PCA spectra above 8 keV to make one bin out of 4 native bins between 8.1 and 16.5 keV, and one bin out of 10 bins between 16.5 and 26.6 keV. This improves slightly the precision on the parameters.

We also checked the data from the High Energy Timing Experiment (HEXTE), the second instrument on-board \rxte. However, as already pointed out in Paper I, no HEXTE data can sensibly be used in our analysis because of the faintness of the source.

The {\tt{PCABACKEST}} software computes the total instrument and cosmic X-ray background. Since \igr\ is a faint source, and lies in the Galactic Ridge, we also corrected the spectra for the Galactic X-ray background (GXB). \citet{Valinia} quantified the diffuse GXB in several regions of the Galactic ridge, including one with Galactic Latitude in the range -1.5--1.5\degr\ and Galactic Longitude in the range -40--40\degr, which corresponds to our source. They used a three component model in the 3--35 keV range: a Raymond-Smith model and a power law, attenuated by interstellar absorption with a column density of $N_{\mathrm{H_{IS}}}$ $\sim$ 1.5 10$^{22}$ cm$^{-2}$. Note that the model absorption measures the Galactic column density through the whole Galaxy in a given direction, so it overestimates the absorption experienced by  \igr\ which is most probably situated between 2 and 6 kpc. As we are interested in the relative variations of the absorption, and since the Galactic absorption in the direction of the source is constant, the precise value of the latter is not needed as it just adds a constant to the absorption intrinsic to the source. 

Moreover, the smallest measurement of the absorption we have near phase 0 with a meaningful error bar is 3.9 $\pm$ 2.3 10$^{22}$ cm$^{-2}$, once the Galactic contribution has been removed. This means that, even if we assume the minimum intrinsic absorption is $\sim$ 1.6 10$^{22}$ cm$^{-2}$, this is still bigger than the Galactic absorption in the direction of the source. So we can reasonably consider the absorption we measure as highly dominated by intrinsic absorption. In the following, we assume that what we measure is the local absorption, which is, hereafter, referred to as \nh. The \rxte/PCA spectra were fitted in XSPEC version 11.3.2, between 3 and 25 keV. A joint \rxte-\inte\ spectrum is shown in Fig. \ref{exemple}, right.

\section{Results}
\label{resultats}

\subsection{Spectral analysis}

We used the ephemeris of \citet{periode2} to fold the spectra, starting at phase 0 when the flux is at a minimum. The analysis of the absorption underwent by the X-ray flux along the orbit shows that phase 0 also corresponds to a minimum of the absorption (see section \ref{model}). Thus, phase 0 corresponds to when the compact object is located between the Earth and the companion star (hereafter inferior conjunction): in this case, the X-ray flux coming from the compact object travels a shorter distance in the companion stellar wind. Phase 0.5 is therefore when the compact object is behind the companion star (superior conjunction). The fact that a minimum flux corresponds to a minimum absorption may be puzzling, but actually these two effects act on two different energy ranges: indeed, the modulation of the ASM flux is seen only in the 5-12 keV range \citep{periode}, where the photoelectric absorption has no influence. Qualitatively, when the neutron star is behind the companion star, we see directly the shock between the stellar wind and the ionized gaz surrounding the compact object, where the high-energy emission is produced. At inferior conjunction, however, the high energy emitting region is partially occulted by the neutron star: thus, this moment appears as a minimum flux in the ASM. 

The phases are calculated at the mid-point of the observations. One observation typically lasts 0.005 phase. The 0.003 day uncertainty in the period determination of 13.552 days by \citet{periode} leads to an uncertainty of 0.02 on the phase determination over the 4 years of observations.

We tested several models in analysing the spectra, starting with phenomenological models. The data were well fitted using an absorbed power law combined with a high energy cut-off. An iron fluorescent line at 6.4 keV is present in the observations where the source is detected at a high enough significance (see Fig. \ref{exemple}, right). Note that the high energy cut-off is only detected in the observations where the source flux is sufficiently high, thus giving enough precision at the high energy end of the spectra. Fig. \ref{exemple}, left, shows the residuals from the fit procedure when we add successively each component of the model. Note that for this observation, we need an additional component to accurately describe the spectrum below 6 keV. The spectrum shows a ``soft excess" feature in this range, which is discussed below.

\begin{figure}
	\flushleft
	\includegraphics[width=8cm]{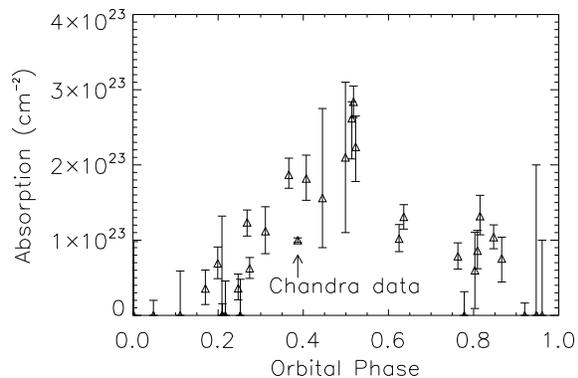}
	\caption{Evolution of the absorbing column density along the orbital phase of the system. Error bars are shown at the 90\% confidence level. ``Chandra data" refers to the value of \nh\ measured by \protect{\citet{optique}}.}
	\label{nH}
\end{figure}

\begin{table*}
%\vbox to 220mm{\vfil Landscape table to go here.
%\caption{} \vfil} 
\flushleft
\caption{}
\label{obss}
\includegraphics[width=16cm]{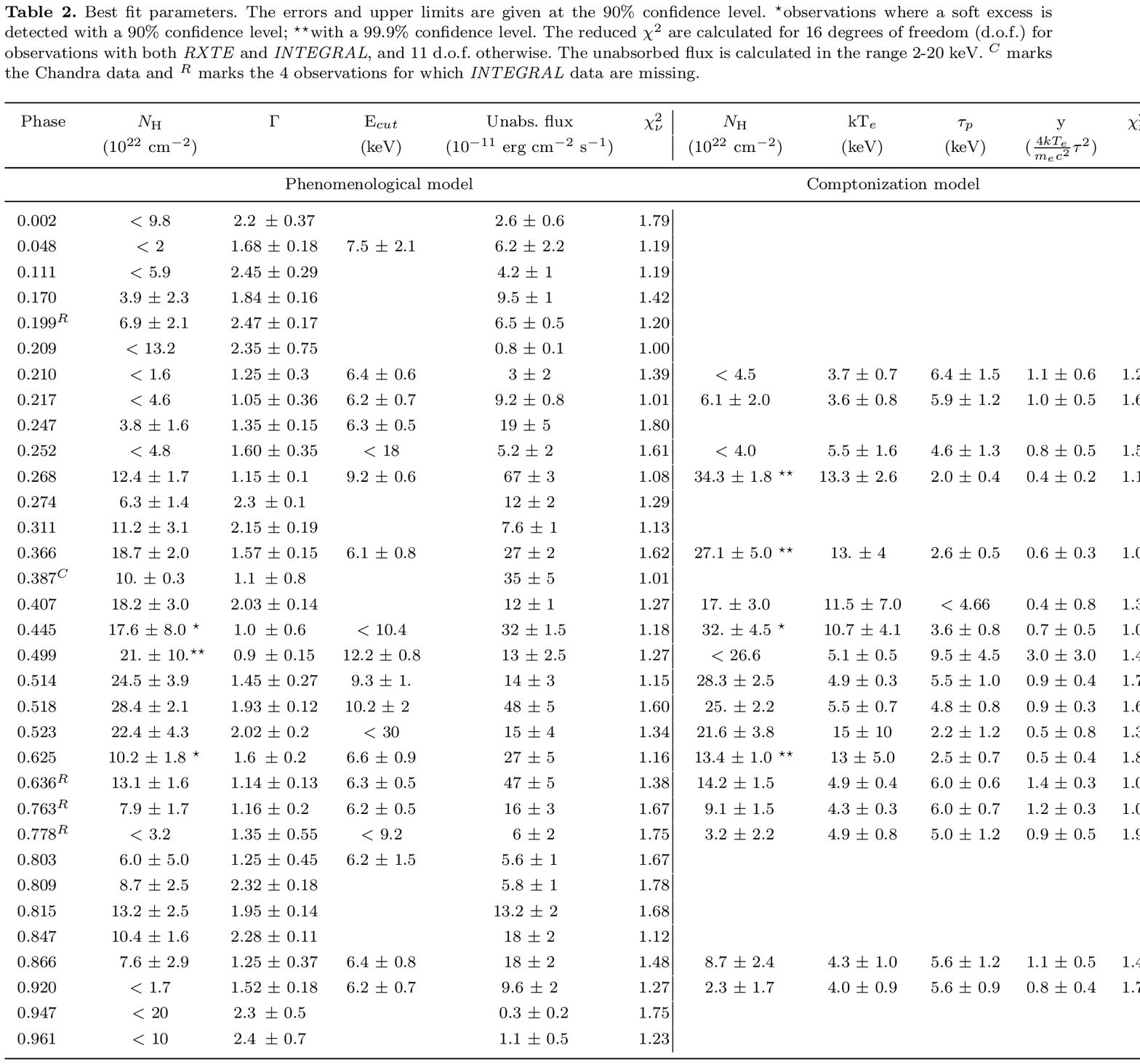}
%\caption{Spectral results}
\end{table*}
\setcounter{table}{2}

\begin{figure*}
	\centering
	\includegraphics[width=8cm]{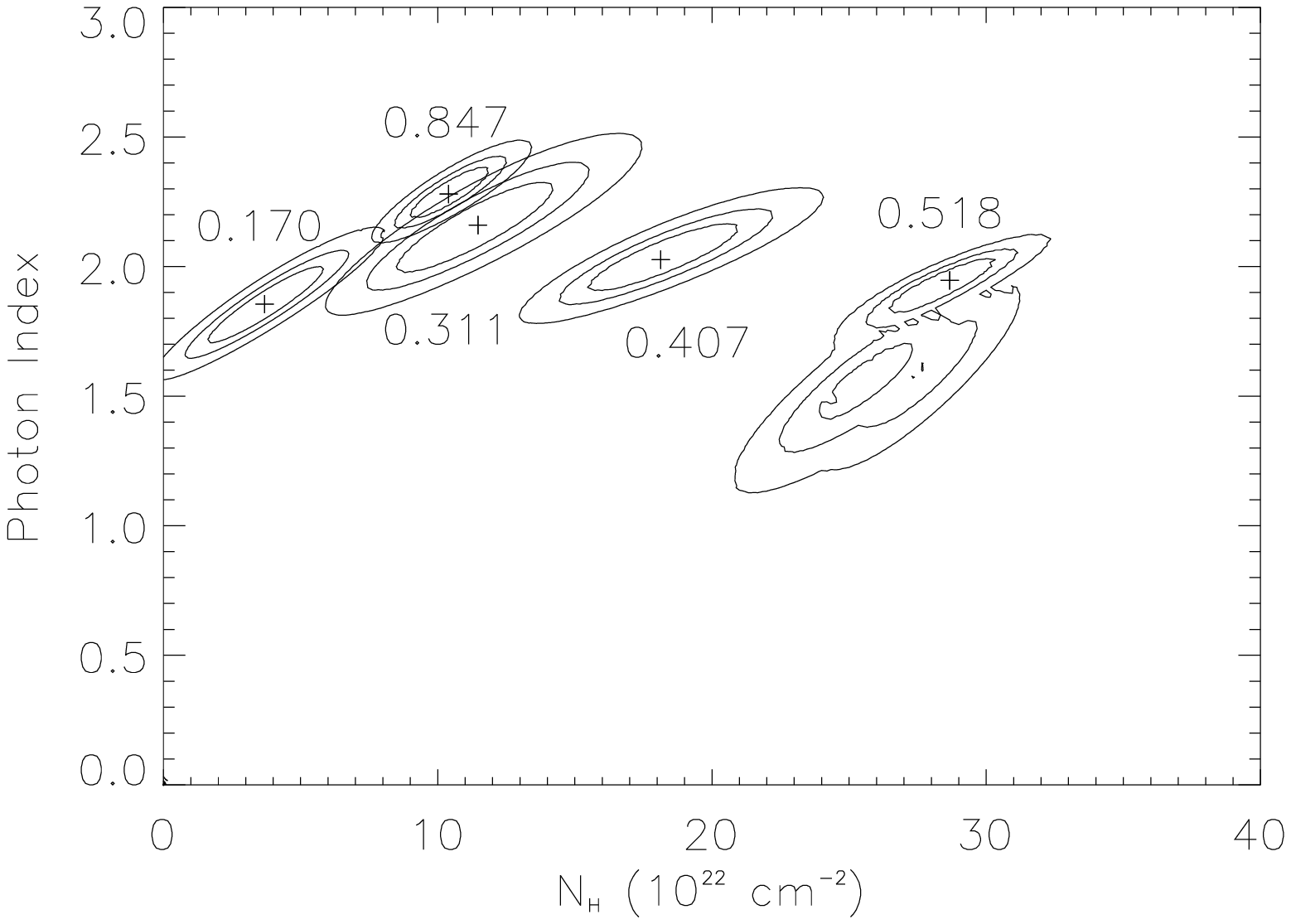}
	\includegraphics[width=8cm]{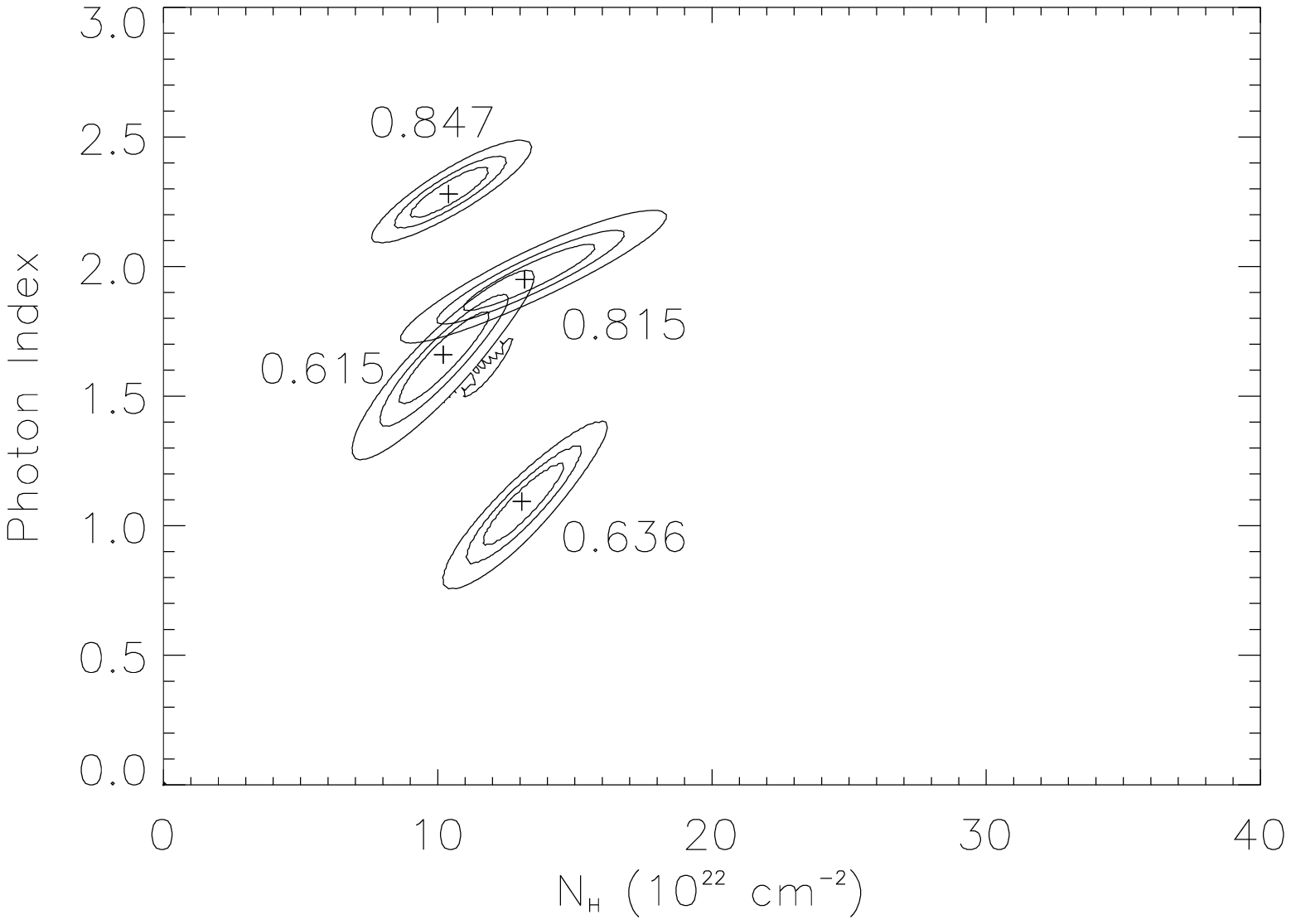}
	\caption{Contour plots of $\chi ^2$ in the \nh-$\Gamma$ plane from fits of eight spectra. The contours correspond to the 68, 90 and 99\% confidence limit, the central cross being the most probable value. The numbers next to the contours are the orbital phase of each observation. \textbf{Left: } Five observations with different \nh\ values and similar $\Gamma$. \textbf{Right: } Four observations with similar \nh\ values and different $\Gamma$.}
	\label{contour}
\end{figure*}

\subsection{Absorption evolution}

The most important feature that we found is a strong evolution of  \nh\ along the orbit (Fig. \ref{nH}). In order to check whether the evolution of \nh\ was genuine or not, we produced power law photon index ($\Gamma$) vs. \nh\ contour plots for all observations. Eight typical observations are reported in Fig. \ref{contour}. These correspond to spectra whose parameters are well constrained. The outer contour represents the 99\% confidence level. The parameters $\Gamma$ and \nh\ are correlated: any variations of \nh\ will have a strong impact on the measure of $\Gamma$ and vice-versa. The elliptical elongation of each individual contour is the result of this link in the fitting process. The contours are clearly distinct from one observation to the next, which clearly indicates that the evolution of \nh\ and $\Gamma$ are genuine. If this were not the case, the contours would overlap. The detailed study of the evolution of \nh\ along the orbital phase is presented in Sec. \ref{model}.

\begin{figure}
	\flushleft
	\includegraphics[width=8.5cm]{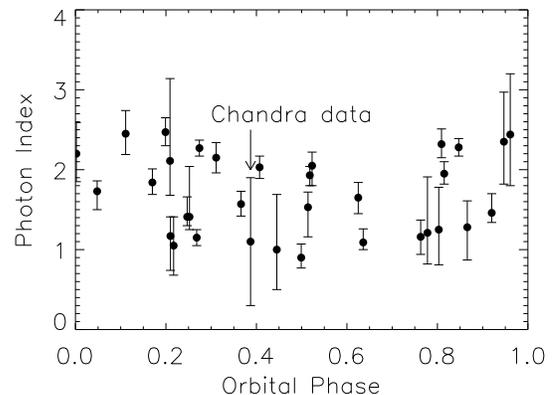}
	\caption{Evolution of the photon index along the orbital phase of the system. Error bars are shown at the 90\% confidence level. ``Chandra data" refers to the value of $\Gamma$ measured by \protect{\citet{optique}}.}
	\label{gamma}
\end{figure}

The absorption measured by \textit{Chandra} is more precisely determined because of the better spectral resolution and the lower energy boundary of this instrument in the soft X-ray range. It is $\sim$ 40\% lower than the general tendency observed by \rxte\ around the same phase. The discrepancy may have two origins. Firstly, there may be a difference of calibration between the two instruments. Secondly, the \textit{Chandra} high energy range is limited to 10 keV, so it gives bad constraints on the index of the power law, as can be seen in Fig. \ref{gamma} (\citet{optique} obtained \nh=1.0 $\pm$ 0.3 10$^{23}$ cm$^{-2}$ and $\Gamma$=1.1 $\pm$ 0.8). Since increasing $\Gamma$ would increase \nh, confidence levels in the \nh-$\Gamma$ plane could increase the value of the error on \nh. Given these uncertainties the value of \nh\ obtained with the Chandra observation was excluded from our wind model fits.

The absorption measured using the Comptonization model differs from the phenomenological model for some observations, especially in the observations where a soft excess is detected, but its overall evolution remains the same. It confirms the reality of this evolution, and the discrepancy is inherent to the differences between the models. Because the phenomenological model fits all the spectra, and in a view to obtain a consistent analysis, we used only the values of \nh\ obtained from the phenomenological model in our studies.

\subsection{Thermal Comptonization}

Since a cut-off power law is usually interpreted as a signature of thermal Comptonization, we replaced this simpler model with a more physical Comptonization model \citep[\texttt{COMPTT,}][]{titarchuk}. In \igr, no stable accretion disc nor thermal component has been detected, so we fixed the disk temperature parameter $kT_0$ at 0.1 keV. The $\chi_\nu^2$ obtained when fitting with this model (modified by photoelectric absorption) were comparable to those of the phenomenological models. Since the parameters are constrained by the high energy end of the spectra, only  observations with a relatively high flux, hence a relatively good signal/noise ratio at high energies, are suitable for the Comptonization model.

\subsection{Soft Excess detection}

In some observations, the $\chi_{\nu}^2$ was far higher than for the other observations, the discrepancy between the model and the data being particularly important at low energies. In those cases, the residuals between the model and the spectrum show that some soft excess is present (see Fig. \ref{exemple}, left). This feature has already been observed in many X-ray binaries \citep[see][for a review]{soft}. We tried to add systematically a black body component at low energies to all our observations, obtaining a better $\chi_{\nu}^2$ for several of them.  In four observations, an F-statistic test indicated values greater than 99.9\% for a true improvement of the fit with the black body, giving a very good confidence in this detection. In one more observation, the test gave a value greater than 90\%. The spectral characteristics of these observations are given in Table \ref{mesures}.

Note that other models, such as a Raymond-Smith model, could also account for the soft excess, but we lack the resolution at low energies to discriminate between the different possibilities, so we kept the simplest one, i.e. the black body model. Besides, since the black body and the absorption models influence the spectra mostly below 5 keV where the PCA coverage and resolution are moderate (only 6 spectral channels cover the range 3--6 keV), it is difficult to separate their respective contributions, and hence we obtain large uncertainties for \nh. Similarly, the black body temperature that we found in the range 0.3--0.6 keV must be treated with caution. Other X-ray binaries have soft excess temperatures in the range 0.1--0.2 keV; a black body with this temperature has weak influence above 3 keV where \rxte\ is sensitive, so the values we obtain should be regarded more as an order of magnitude than absolute values.

Adding a black body at low energy increases the uncertainties as it adds a degeneracy in the fitting process: increasing the black body flux compensates for a decrease in the absorption level. Therefore we verified, by drawing confidence levels in the kT-\nh\ plane, that the black body temperature is clearly determined. Given the high uncertainties on this latter parameter, however, the contours of several observations in which the soft excess is detected overlap. We, therefore, cannot conclude on any evolution and/or relation of the temperature with orbital phase. 

\begin{table}
\caption{Spectral characteristics of the five soft excess observations.}
\label{mesures}
	\begin{minipage}{83mm}

		\begin{tabular}{cccccc}
			\hline
			Phase &  kT & \nh \footnote{Phenomenological model.} &  $\Gamma$    & $\chi_{\nu}^2$ \footnote{Calculated for 15 d.o.f. for the phenomenological model. The Comptonization model gives $\chi_{\nu}^2$ values with differences of $\pm$ 0.15.}& y\footnote{Kompaneets parameter obtained from the Comptonization model.} \\

			\hline
			0.268 &  0.27 $\pm$ 0.02    & 12.4\z  $\pm$ 1.7 & 1.15 $\pm$ 0.1 & 1.08 & 0.4 \\ 
			0.366 &  0.27 $\pm$ 0.10    & 18.7\z  $\pm$ 2.0 & 1.57 $\pm$ 0.2 & 1.62 & 0.6 \\
			0.445 &  0.22 $\pm$ 0.04    & 17.6\z  $\pm$ 8.0 & 1.00 $\pm$ 0.6 & 1.18 & 0.7 \\
			0.499 &  0.46 $\pm$ 0.25    & 21.\z\z  $\pm$ 10. & 0.90 $\pm$ 0.2 & 1.27 & 3.0 \\
			0.625 &  0.27 $\pm$ 0.08    & 10.2\z  $\pm$ 1.8 & 1.60 $\pm$ 0.2 & 1.16 & 0.5 \\
			\hline
		\end{tabular}
	\end{minipage}
\end{table}

\section{Discussion}
\label{discut}

\subsection{Comparison with previous work}

A comprehensive analysis of the very first X-ray observations of \igr\ is reported in Paper I. The goal of this study was to try to understand the nature of this, then newly discovered source. At that time only a set of three \rxte\ observations and a limited number of simultaneous JEM-X and ISGRI data were available. Instead of performing an orbital phase resolved study, and in order to improve the signal to noise ratio of the \inte\ data, we performed a time-resolved spectroscopy based on the 20--40 keV luminosity level of the source. As a result it is very likely that we  mixed the spectra extracted at different orbital positions. This is particularly obvious if one tries to compare the results obtained in paper I and the ones we report here. Although qualitatively they match quite well, i.e. the spectral models are similar, and the sets of parameters are compatible, none of the ``states" identified through the \inte\ data in Paper I were observed here.

First of all it has to be noted that during the fitting of the \inte\ spectra in Paper I, the value of \nh\ was frozen to that found in \citet{Atel}, i.e. $6\times 10^{22}$ cm$^{-2}$, while it is clear from the present study that \nh\ is tightly linked to the orbital phase. Secondly, the procedure of summing spectra taken at similar fluxes in Paper 1 may have allowed us to pick up peculiar phases of accretion, as e.g. flares due to enhanced accretion. The analysis reported here focuses more on a global and ``normal" behaviour of \igr. Finally, since the analysis of the \inte\ data in Paper I was based on the detection by ISGRI, it is clearly biased towards states that are bright at hard X-rays. Our orbital-phase dependent analysis showed us that there was not necessarily a connection between the level of hard X-rays and the position on the orbital phase. Note also that although \igr\ and GRS~1915+105 are 1.1\degr\ apart, some confusion is still possible in the JEM-X data.

Two of the \rxte\ observations presented here were already analyzed in Paper I. They respectively correspond to phase 0.763 (Obs. 3 in Paper I) and 0.815 (Obs. 2 in Paper I). One can immediately see that the results obtained in both analyses clearly differ, in particular in the value of the absorption (and consequently the unabsorbed fluxes). There are two reasons that are the very probable origin of the discrepancies. First, the analysis presented in Paper I was made with old versions of the software and history file that could cause errors on the background estimation of up to 10\%. This problem has been solved since then. The second origin, which is probably the main reason for the differences, is that in Paper I the Galactic background was not taken into account in correcting the data. This has a strong effect at low energies, and hence influences highly on the determination of the absorption.

Qualitatively, however, the spectral behaviour of \igr\ is confirmed. We confirm that an iron K$\alpha$ fluorescence line is present in most of the spectra. This line is intrinsic to the source, since if it were due to the Galactic background, we would expect its flux to be roughly constant, which is not the case here. We measure upper limits on its width in the range 300-1000 eV, depending on the line flux. This could be indicative of a narrow line, rather than a broad line, but actually we are mainly limited by the instrumental spectral resolution.

\subsection{Spectral evolution}
\subsubsection{Comptonization model}

Our first spectral analysis showed that a simple cut-off power law model could accurately fit all the spectra. Therefore we applied a thermal Comptonization model, as this is the usual physical model which explains this feature. We obtained the parameters kT$_e$ and $\tau$ reported in Table \ref{obss}, which are consistent with similar observations of other HMXBs and absorbed sources detected by \inte\ \citep[e.g. IGR J16320-475,][]{IGR_1632}. The Comptonization parameter $y = \frac{4kT_e}{m_e c^2} \tau^2$ (also known as the Kompaneets parameter) determines the efficiency of the Comptonization process, and thus the shape of the spectrum \citep{titarchuk}. In our case, this parameter is $\approx$ 0.5--1, indicating a moderately efficient process. It corresponds to a rather low accretion rate, which is consistent with the supposed nature of the system: wind-fed accretion is less efficient than Roche lobe overflow. Our estimation of this mass-loss rate (Sec. \ref{fin}) in the range 4--7$\times$10$^{-8}$ M$_{\sun}$/year corroborates the B-type nature of the companion. Note that it has no significant evolution along the orbital phase, showing that the emission process is remarkably stable along the orbit. This suggests a rather stable accretion rate, and therefore would tend to favour a circular orbit for the system.

\subsubsection{Phenomenological model}
 
Several observations show a photon index $\Gamma \sim 1$, for instance 1.15 $\pm$ 0.1 at phases 0.268 and 0.636, or 1.05 $\pm$ 0.36 at phase 0.217. This is consistent with a neutron star nature of the compact object, as these hard spectra are typical of saturated Comptonization found in binary systems containing a neutron star \citep{emission}.

Having identified the emission process as thermal Comptonization, we returned to the phenomenological model, which allows all the spectra to be fitted and permits a true study of the absorption evolution to be done. Before studying the evolution of \nh, we can draw some conclusions from the other parameters.

The unabsorbed flux undergoes strong variations that are not correlated with the orbital phase. The variations probably arise from the details of the accretion process. The stellar wind is not perfectly homogeneous and the accretion onto the neutron star, following the magnetic field line to the magnetic poles, is a highly unstable and complicated process \citep[see e.g.][and references therein, for a review of the emission processes]{outer}. Since our observations are spread over several years, it is not surprising that the source flux is highly variable. However, if the orbit of the neutron star were eccentric, when the compact object approaches the companion star where the stellar wind density is higher, we would see an increase in the source flux. Such evolution is marginally visible, the unabsorbed flux being smaller between phases 0.9 and 0.2. However as this is only marginal, we interpret it as another point in favour of a circular orbit for the system.

The photon index of the power-law component undergoes strong variations as well (Fig. \ref{gamma}). These variations are not correlated with the flux evolution or the orbital phase, which means that this is probably just another effect of the high variability of the accretion process.

\subsubsection{Origin of the Soft Excess}
During our study, some spectra exhibited an excess in the soft X-ray part of the spectra, which we modeled by adding a black body component to the model. Following \citet{soft}, we can look for possible mechanisms explaining this feature. Using the parameters listed in Table \ref{param}, the orbital separation between the two objects $a$ ($a = \left[\frac{G M _{\star} P_{orb}^2}{4 \pi^ 2}\right] ^{1/3}$) is in the range 51-74 R$_{\sun}$. If we assume a mass for the neutron star of $\sim$ 2 M$_{\sun}$, then the Lagrange point L$_1$ is located at a distance from the companion star in the range 34-56 R$_{\sun}$. Thus, it is very unlikely that the companion star overflows its Roche lobe, which is confirmed by the fact that we do not observe any accretion disk. This rules out explanations for the soft excess feature involving an accretion disk or a gas stream flowing from the companion to the compact object \citep{stream}. We can also rule out soft emission from the accretion column, as it would imply soft pulses in the spectra.

According to \citet{soft}, this leaves two possible processes which involve the immediate surroundings of the compact object, characterized by the influence of the compact object X-ray emission. The stellar wind of massive stars is mostly accelerated by bound-bound transitions of atoms. The neutron star, due to its high energy emission, ionizes the surrounding material, so the already ionized gas around the neutron star is no longer accelerated by the stellar radiation field \citep{vent_ralenti}. When the compact object moves along its orbit, the hot gas will gradually be overtaken by the stellar wind. This will lead to the formation of a ``tail" trailing the neutron star.

The stellar wind collides with this tail, transforming a fraction of its kinetic energy into X-ray emission. This results in a shock located between the compact object and the secondary. According to \citet{soft}, this could explain the soft excess of faint sources ($L_X \la 10^{36}$ erg s$^{-1}$), which is the case of \igr\ ($L_X \approx 3. 10^{35}$ erg s$^{-1}$ for a distance of $\sim$ 5 kpc and a characteristic X-ray flux of $\sim$ 10$^{-10}$ erg cm$^{-2}$ s$^{-1}$). Finally, the last process involves the ``tail" itself: the diffuse cloud around the neutron star may reprocess the hard X-rays coming from the compact object.

However, to differentiate between emission produced in collisionally or photoionized plasma, high resolution grating spectra are required, which is unfortunately beyond the possibilities of \rxte. Still, the fact that the soft excess was only seen around the superior conjunction may be of some importance. Until now, only a handful of soft excess detections have been reported and among them two detections, reported for 4U 1700-37 by \citet{orbite} and for Centaurus X-3 by \citet{burderi}, both sources for which the orbital position is known, fall during phase 0.5-0.75. However, this is a too small sample to draw any definite conclusion.

\begin{figure*}
	\begin{center}
	\includegraphics[width=8cm]{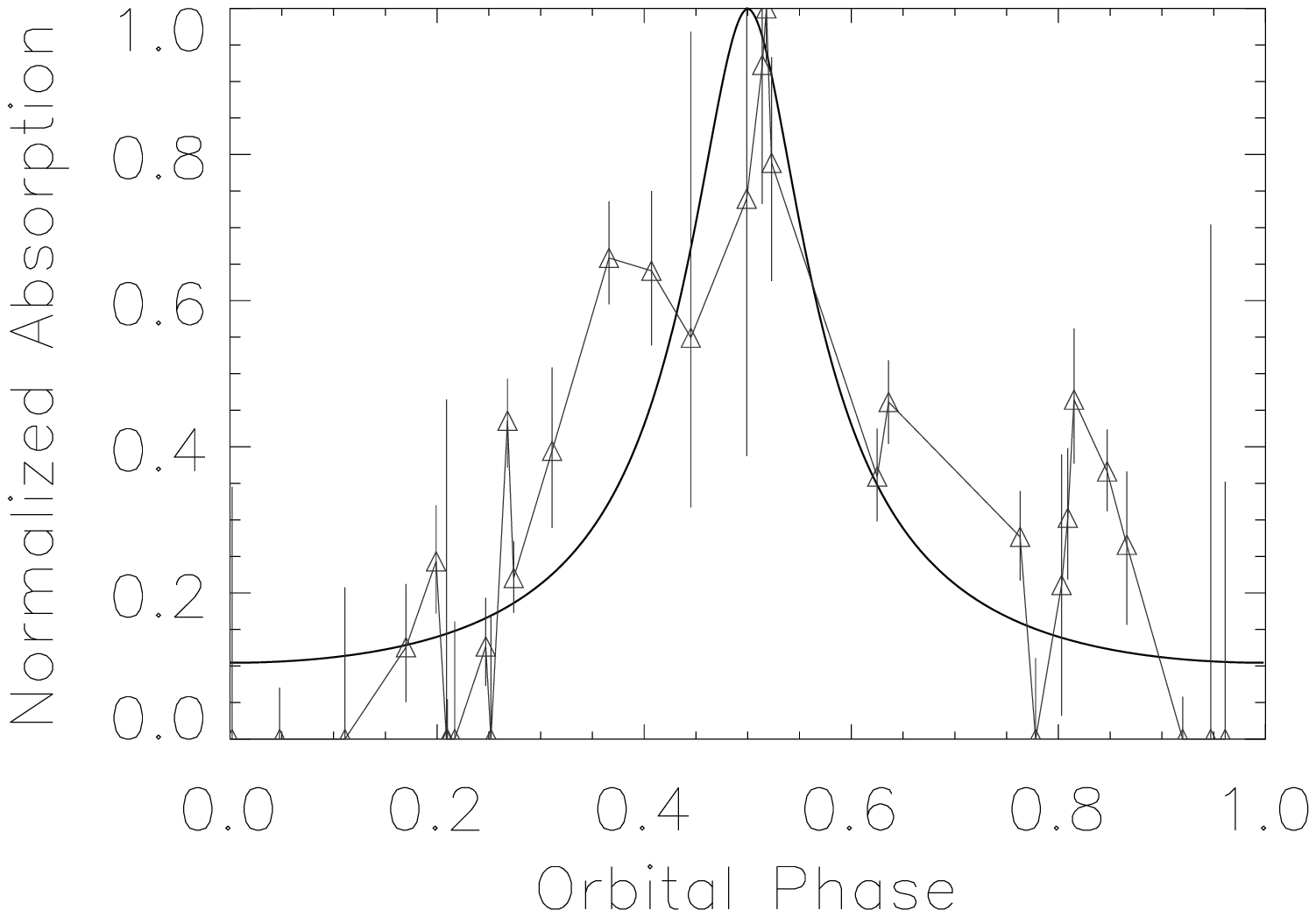}
	\includegraphics[width=8cm]{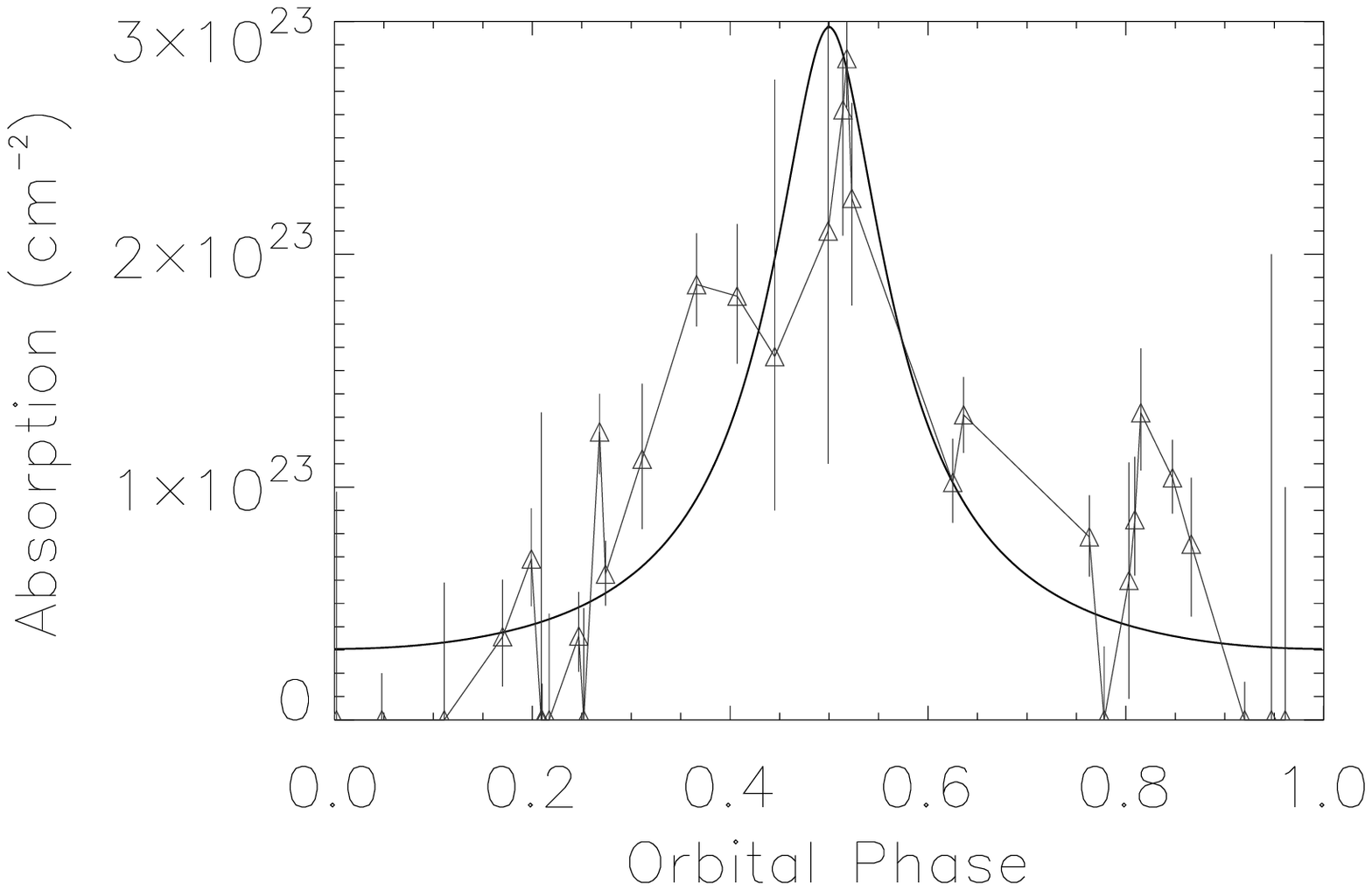}
	\caption{Model absorption (continuous line) and experimental absorption ($\triangle$ symbols), for the stellar model with parameters M$_{\star}$ = 20M$_{\sun}$ , R$_{\star}$ = 21R$_{\sun}$ and $\beta$ = 0.5. {\bf{Left :}}, result obtained with data and model normalised ($\chi^2_{\nu}=1.440$). {\bf{Right :}} Same as left, but without normalisation ($\chi^2_{\nu}=1.432$).}
	\label{meilleur}
	\end{center}
\end{figure*}

\subsection{Wind model}
\label{model}
\subsubsection{Description of the model}
Since the orbit of the X-ray source is probably almost circular, one possible explanation for the observed change in \nh\ relies on an inclined orbit for the system. Qualitatively, when the neutron star is behind its companion its light has to travel a longer distance in the stellar wind, so the absorption in the soft X-rays increases. Therefore, at the superior conjunction we should observe a maximum in the absorption, which is indeed the case (Fig. \ref{nH}). Following this hypothesis, we used a simple wind model in order to constrain several parameters for the system, following the method already used by \citet{inclinaison}. A B-type star emits a strong stellar wind, usually taken to be stationary and spherically symmetric \citep[e.g.][]{castor}. Its velocity is given by a $\beta$-law:

\begin{eqnarray}
V (r) = V_0 + (V_{\infty} - V_0) \left(1 - \frac{R_{\star}}{r} \right) ^{\beta} 
\end{eqnarray}

where $r$ is the distance from the companion star, $R_{\star}$ its radius, $V_0$ the wind velocity at the photosphere, and $V_{\infty}$ the terminal wind velocity. Observations show that $V_{\infty}$ typically lies in the range 1000--1500 km~s$^{-1}$ for a B1 star \citep{vinf}. For early-type stars, $\beta$ is in the range 0.7--1.2 \citep[e.g.][]{O_massloss}. We have $V_0 \ll V_{\infty}$, so for our purposes we can take:

\begin{eqnarray}
V (r) \simeq V_{\infty} \left(1 - \frac{R_{\star}}{r} \right) ^{\beta},
\end{eqnarray}

\begin{figure}
	\begin{center}
	\includegraphics[width=4cm]{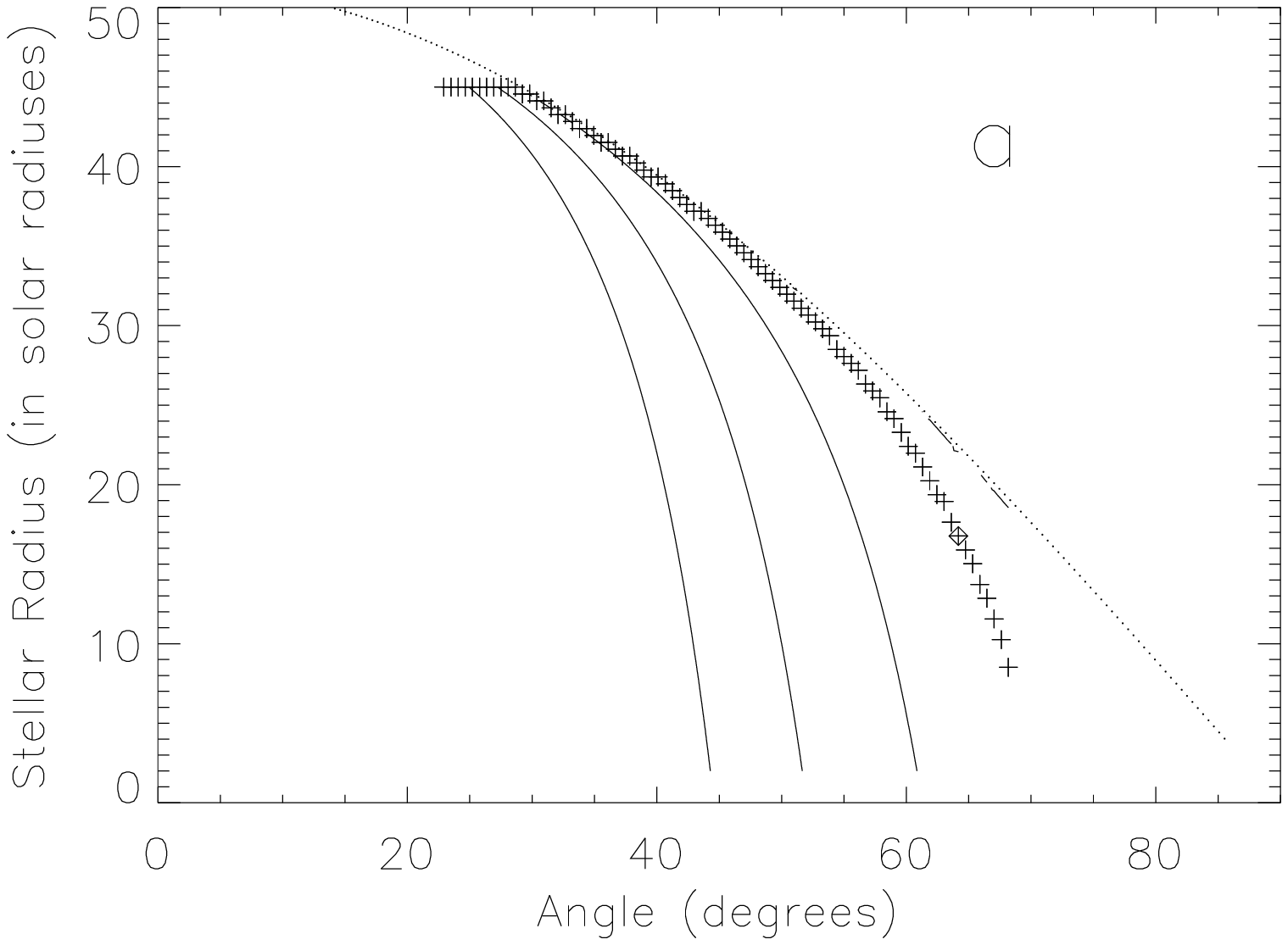}
	\includegraphics[width=4cm]{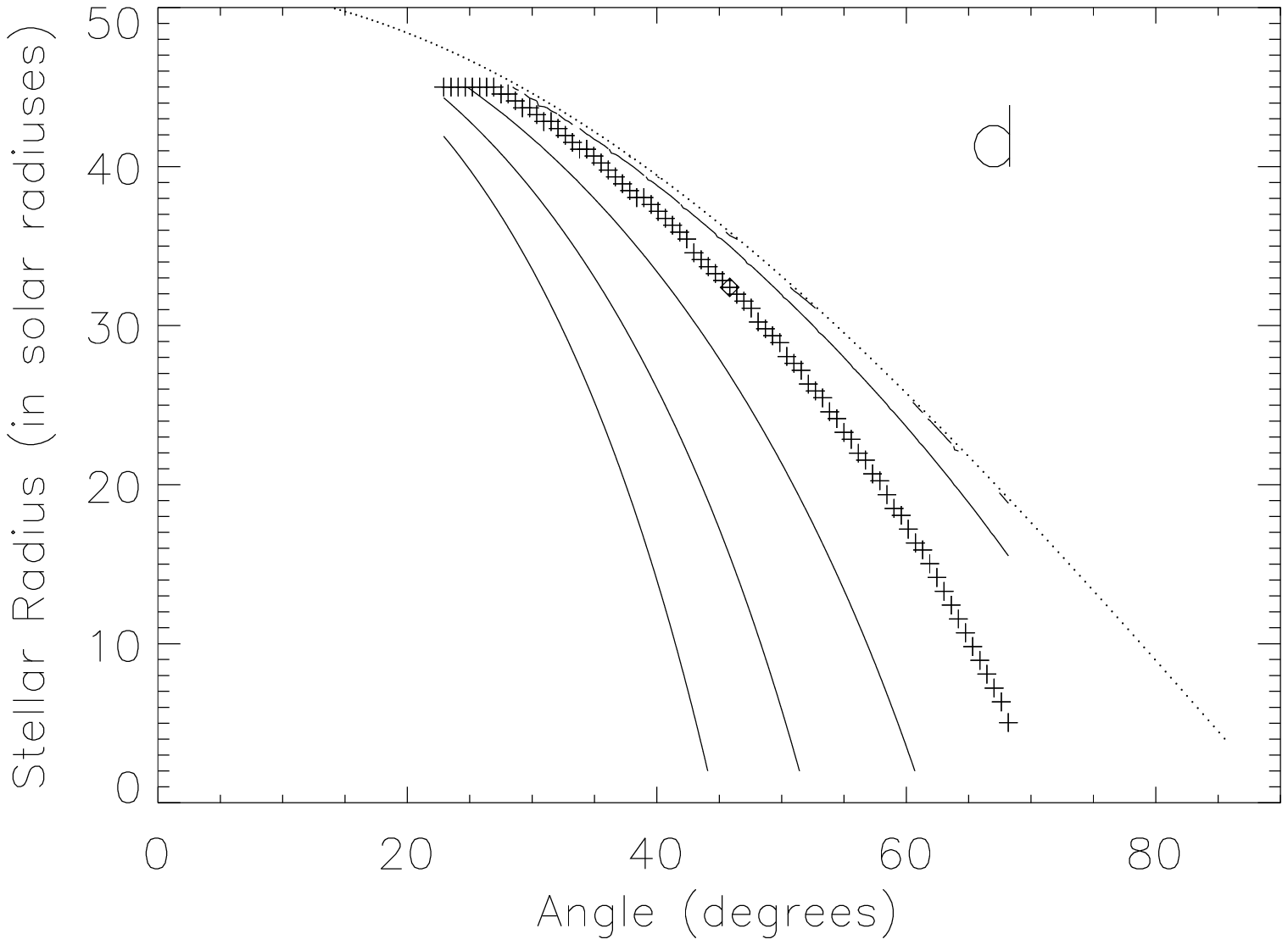}
	\includegraphics[width=4cm]{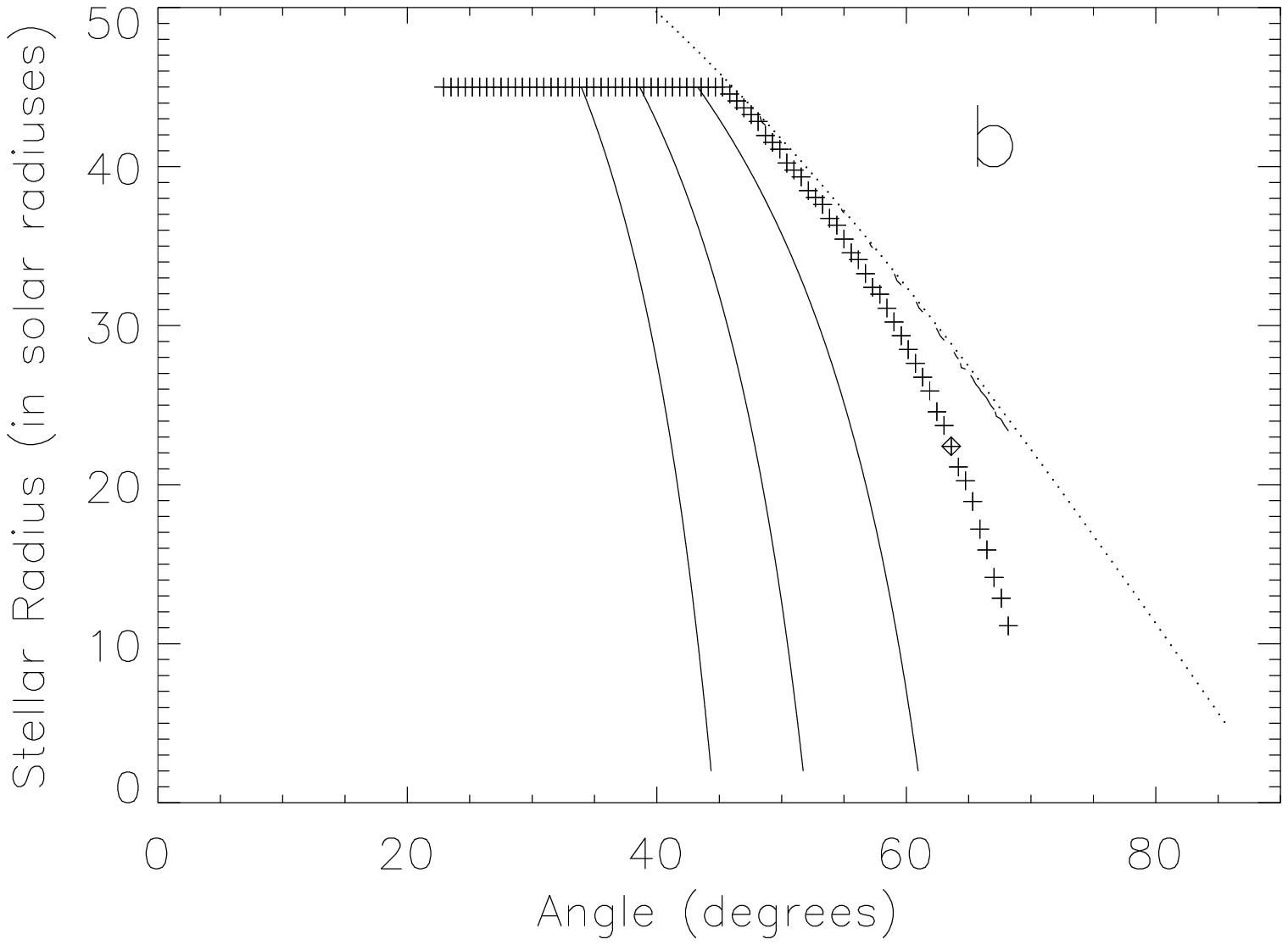}
	\includegraphics[width=4cm]{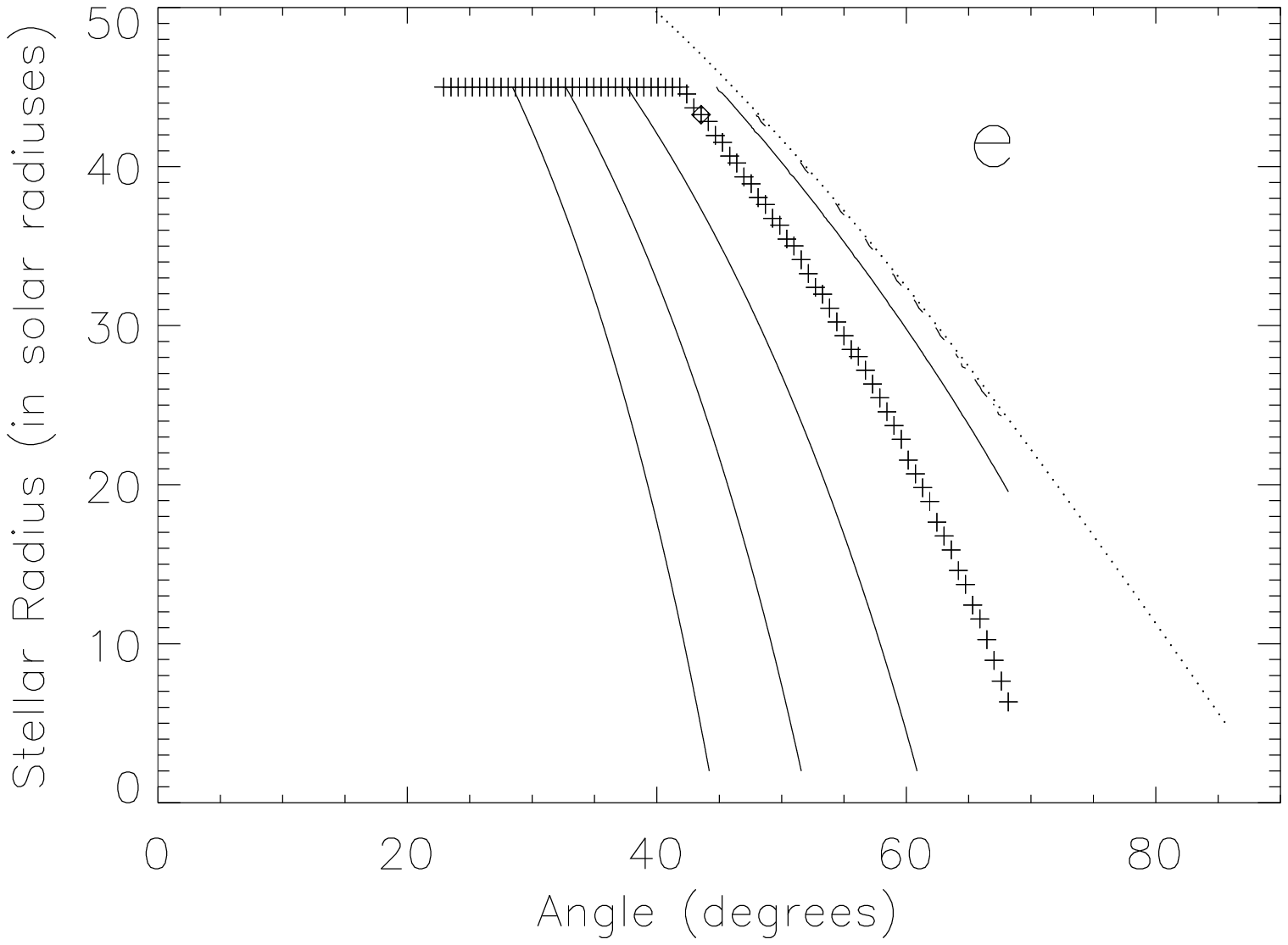}
	\includegraphics[width=4cm]{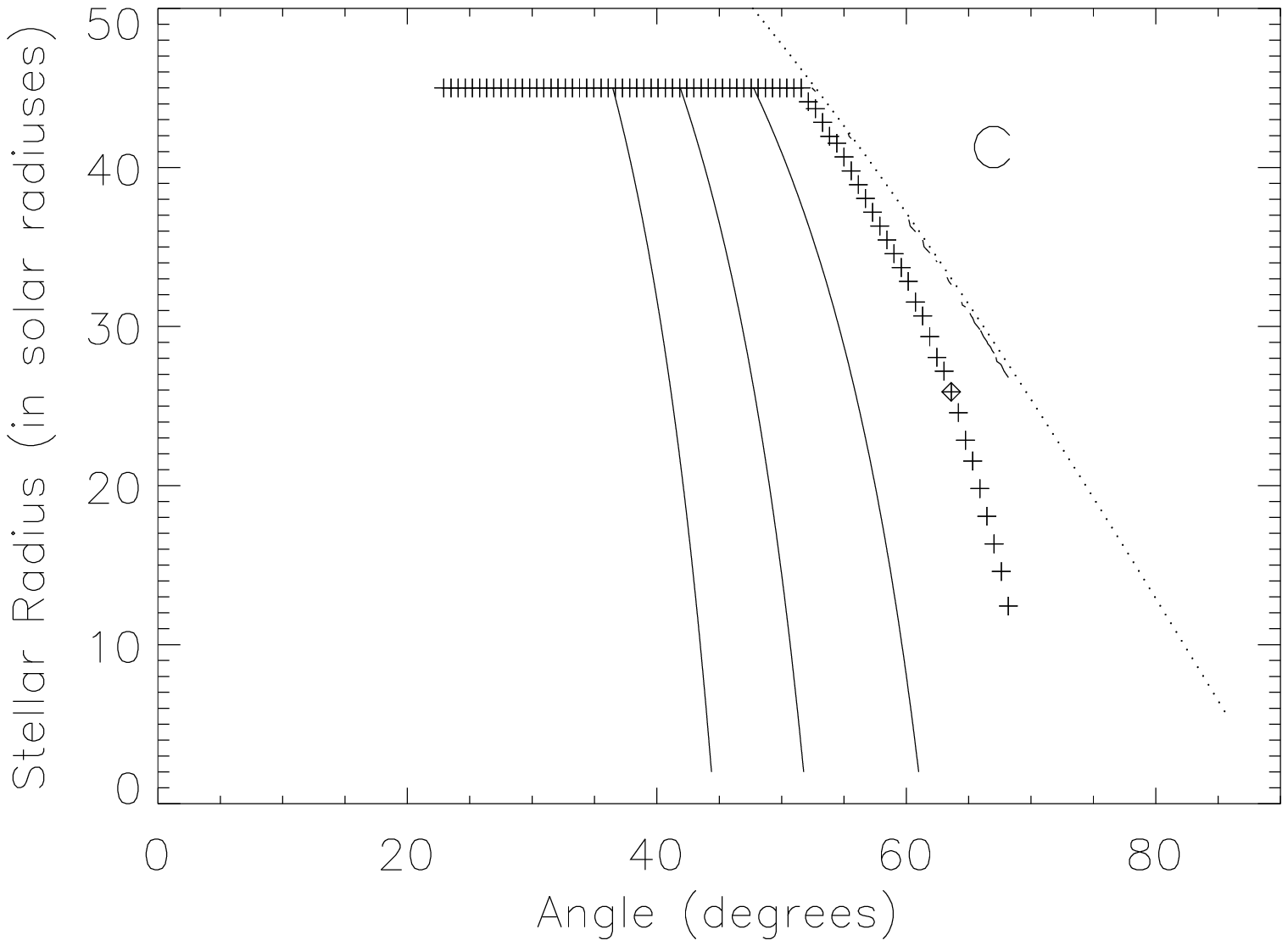}
	\includegraphics[width=4cm]{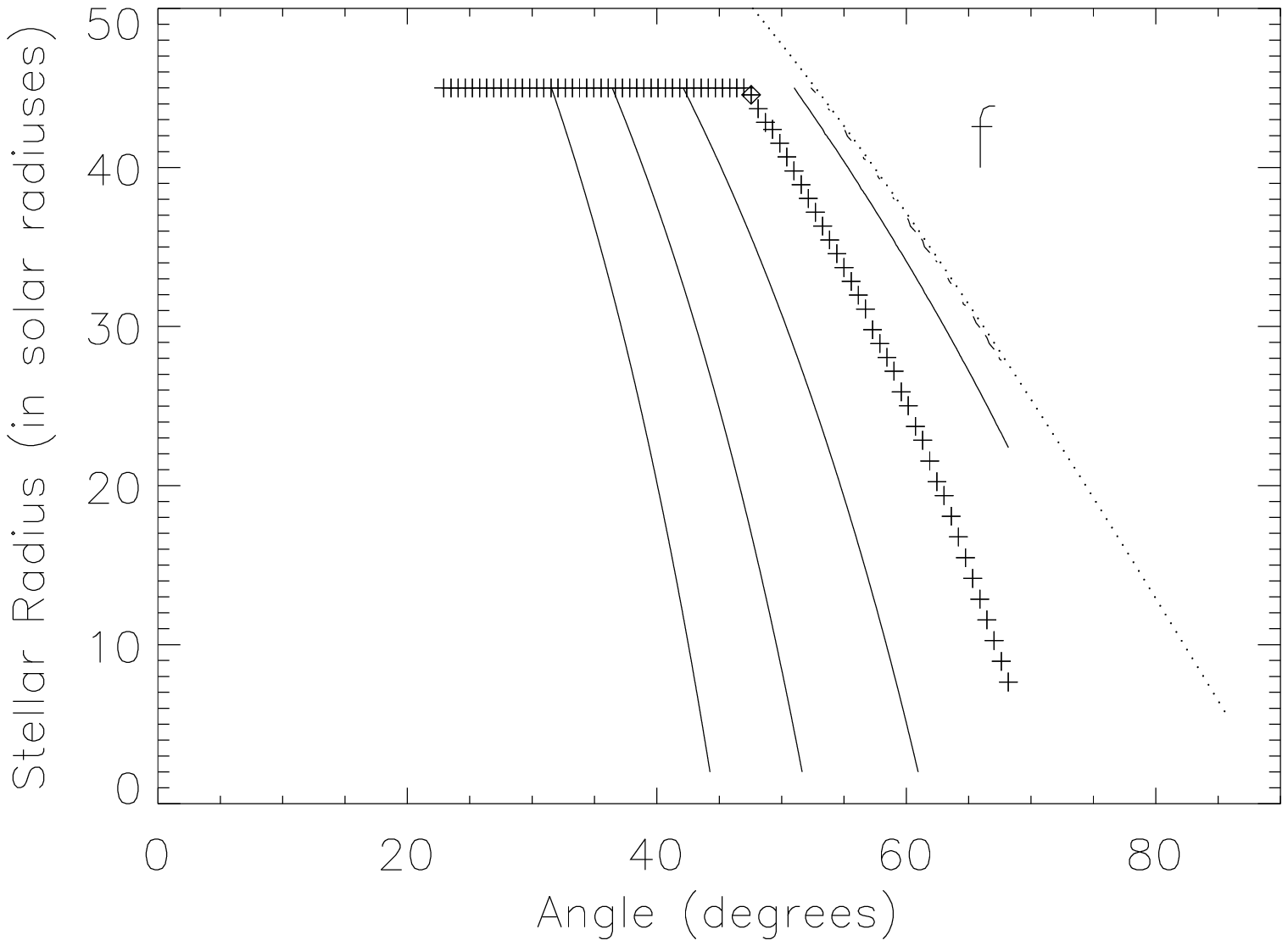}
	\end{center}
	\caption{Most probable orbital inclination, as a function of the companion radius, for various $M_{\star}$ and $\beta$. {\bf{a :}} $\beta$=0.5, $M_{\star}$=10$M_{\sun}$. {\bf{b :}} $\beta$=0.5, $M_{\star}$=20$M_{\sun}$. {\bf{c :}} $\beta$=0.5, $M_{\star}$=30$M_{\sun}$. {\bf{d :}} $\beta$=1, $M_{\star}$=10$M_{\sun}$. {\bf{e :}} $\beta$=1, $M_{\star}$=20$M_{\sun}$. {\bf{f :}} $\beta$=1, $M_{\star}$=30$M_{\sun}$. The + symbols correspond to the most probable orbital inclination, confidence contours are drawn at the 25, 68 and 90\% confidence levels. The dotted line is the ``eclipse limit": points above this line are excluded as it would imply an eclipse of the neutron star by its companion, which is not observed.}
	\label{simm}
\end{figure}

\begin{figure*}
 \centering
 \begin{minipage}{140mm}
	\includegraphics[width=9cm]{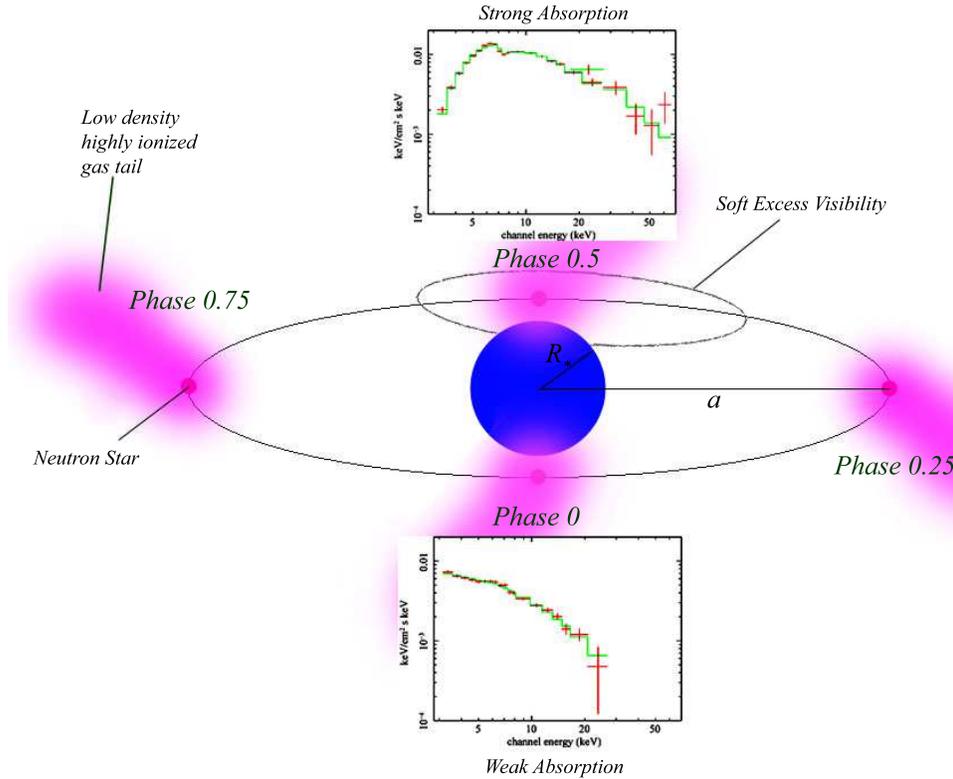}
	\caption{Diagram of \igr\ as it could be seen from the Earth. The orbital inclination of the system is taken to be $\sim65$\degr.}
	\label{schema_recap}
	\end{minipage}
\end{figure*}

The equation of mass conservation then gives

\begin{eqnarray}
 \dot{M_{\star}} = 4 \pi r^2 V (r) \rho(r)
\end{eqnarray}

where $\dot{M_{\star}}$ is the mass-loss rate of the star, and $\rho(r)$ the density. The number density $n(r)$ of hydrogen atoms is in turn given by $\rho(r) = \mu m_H n(r)$, where $\mu$ is the mean atomic weight of the particles expressed in units of the hydrogen atom mass $m_H$ ($\mu = 0.602$ for solar composition material). The instantaneous equivalent absorbing column density of hydrogen $N_H$ between the neutron star and the observer is given by:

\begin{eqnarray}
N_H & = &  N_{H_{ISM}} + \int^{\infty}_{0} n[r(s)]ds \nonumber \\
    & = &  N_{H_{ISM}} + \frac{\dot{M_{\star}}}{4 \pi \mu m_H V_{\infty}} \int^{\infty}_{0} \frac{ds}{r^2 (1 - \frac{R_{\star}}{r}) ^{\beta}} 
\end{eqnarray}

where $N_{H_{ISM}}$ is the contribution from the interstellar medium, and {\it {s}} is the distance along the line joining the neutron star and the observer. $s$ is given by

\begin{eqnarray}
r^2 = a^2 + s^2 - 2 a s cos(\psi)
\end{eqnarray}

where $\psi$ is the angle subtended at the neutron star between the radial and the observer directions, and $a$ is the orbital separation. Finally, for a circular orbit with orbital inclination $i$, the angle $\psi$ is related to the inclination angle and the orbital phase $\phi$ by

\begin{eqnarray}
cos(\psi) = -sin i \hspace{4pt} cos(\phi(t) - \phi(\tau_{90}))
\end{eqnarray}

where $\tau_{90}$ is the instant of the superior conjunction. Since no eclipse is visible in the system's light curve, we can restrict the inclination angle to

\begin{eqnarray}
i < arccos(\frac{R_{\star}}{a})
\end{eqnarray}

\subsubsection{Parameter fitting}

\label{fin}

Our model needs four parameters, $R_{\star}$, $M_{\star}$, $\beta$ and the ratio $\dot{M}_{\star}$/$V_{\infty}$, and computes the most probable orbital inclination $i$. We first verified the consistency of the ephemeris. We fixed every parameter to the typical values listed in Table \ref{param}, and let the program adjust the maximum of the absorption for several $i$. We obtained maxima between phases 0.49 and 0.5, thus 
confirming the value of $\tau_{90}$. Then, fixing the maximum to phase 0.5, we left free the inclination $i$ and the stellar radius $R_{\star}$, set $M_{\star}$ to 10, 20 and 30 $M_{\sun}$, $\beta$ to 0.5 and 1, and the ratio $\dot{M}_{\star}$/$V_{\infty}$ normalized to the data.

\begin{table}
\begin{minipage}{83mm}
	\caption{Typical parameter values for a B1I star.}
	\label{param}
		\begin{tabular}{ccc}
			\hline
			\hline
			Parameter & value \\
			\hline
			$R_{\star}$       & 10 - 30 $R_{\sun}$                       \\
			$M_{\star}$       & 10 - 30 $M_{\sun}$                       \\
			$\dot{M}_{\star}$ & 10$^{-6} M_{\sun}/ y$ \footnote{\cite{étoiles}}  \\
			$V_{\infty}$      & $\approx$ 1000 km~s$^{-1}$ \footnote{\cite{orbite}} \\
			$\beta$           & 0.5-1                                    \\
			\hline
		\end{tabular}
\end{minipage}
\end{table}

The results are shown in Fig. \ref{simm}. The outer contour lines correspond to the 90\% confidence levels. The two lines at 2 $R_{\sun}$ and 45 $R_{\sun}$ are artifacts caused by the artificial constraint of $R_{\star}$ between these two values. For the most probable stellar radius of 21 $R_{\sun}$, the lower inclination limit is constrained between 38\degr\ and 43\degr, with $\chi^2$ being at a minimum in the range 60--67\degr. Fig. \ref{meilleur}, left, shows the best-fit model against the experimental normalised data, with a good agreement.

We can also determine the ratio $K = \frac{\dot{M}_{\star}}{4 \pi \mu m_H V_{\infty}}$. Using the values given in Table \ref{param}, the predicted value of $K$ is $K_0$ = 5.0 10$^{34}$ atoms~cm$^{-1}$. An estimate of the experimental value of $K$ can be obtained using the best model, whose parameters are M$_{\star}$ = 20M$_{\sun}$ , R$_{\star}$ = 21R$_{\sun}$ and $\beta$ = 0.5. We let the program adjust the normalisation of the model, and obtain $K$ = 0.045 $K_0$ (Fig. \ref{meilleur}, right). The observations predict $V_{\infty}$ in the range 1000--1500 km~s$^{-1}$ for a B1I star \citep{vinf}, which give $\dot{M}_{\star}$ in the range 4--7$\times$10$^{-8}$ M$_{\sun}$/year. This is consistent with the expected value, and retrospectively corroborates the model. However, the wind model used in this analysis is really simple, it does not take into account e.g. a clumping of the wind, since our data are not sufficiently precise to allow a refinement the model. Therefore our numerical results should remain mostly indicative of the evolution of the system.

\section{Conclusions: simple representation of the system}

Fig. \ref{schema_recap} summarizes schematically our main results. We found that photoelectric absorption highly correlates with the orbital phase of the system. Using a simple stellar wind model we found a rather high orbital inclination, $\sim65$\degr. We have detected a soft excess in some observations, just before the superior conjunction, in the area indicated on the figure. This may be explained by a cloud of highly ionised gas surrounding the neutron star. Because of its ionisation, this gas is less accelerated by the stellar wind and, while being gradually overtaken by the wind, tends to form a ``tail". For typical parameters for the secondary star, the neutron star velocity $V_{NS}$ on its orbit is $\sim$100 km s$^{-1}$ and the wind velocity at the position of the orbit is $\sim$0.8 $V_{\infty}$. Therefore, as the diagram shows, the angle between the ``tail" and a vector normal to the orbit is $\theta = arcsin(\frac{V_{NS}}{0.8V_{\infty}}) \approx 10\degr$. This tail could scatter the hard X-ray emission from the compact object and thus explain the soft excess feature. However, these results remain qualitative, as a precise determination of the geometry of the stellar wind would require a better sensitivity in the soft X-ray range, which could be achieved for instance with the use of the \textit{XMM} or \textit{Chandra} satellites.

The study of X-ray binaries is challenging since it is often difficult if not impossible to identify their visible and infrared counterparts. Even if an infrared counterpart were observed, the distance to some systems prohibits the measurement of the orbital characteristics. Our study shows that X-ray observations can overcome these limitations and produce very precise inferences. The \rxte\ and \inte\ observations of \igr\ have led to good measurements of the orbital period of the system and constraints on its inclination angle. 

Moreover, we can use the compact object to probe the stellar wind of the companion. In the case of \igr, we diagnosed the type of the companion (supergiant O or B), the wind density and its structure around the neutron star. More precise observations could lead to constraints on the mass and radius of the companion, and better constraints on the stellar wind. This allows the study of new \inte\ sources, either distant or highly absorbed, and ultimately the determination of new useful data for X-ray binary evolution scenarios.

\section*{Acknowledgments}
We thank T. Foglizzo for useful discussions about the soft excess interpretation. We are grateful to an anonymous referee for a careful reading of the manuscript and constructive comments that significantly improved this paper. The authors warmly thank the \rxte\ and \inte\ planners for having scheduled the observations in simultaneity. DCH gratefully acknowledges a Fellowship from the Academy of Finland. This research has made use of data obtained through the High Energy Astrophysics Science Archive Research Center and quick-look results provided by the ASM/RXTE team. Based on observations with INTEGRAL, an ESA project with instruments and science data centre funded by ESA member states (especially the PI countries: Denmark, France, Germany, Italy, Switzerland, Spain), Czech Republic and Poland, and with the participation of Russia and the USA.
%\bibliographystyle{astron}
%\bibliography{1914}

\label{lastpage}

\end{document}